\documentclass[letterpaper, 10 pt, conference]{ieeeconf} 
\IEEEoverridecommandlockouts
\usepackage{graphicx}
\usepackage{xcolor}
\usepackage{algorithm} 
\usepackage{algpseudocode} 
\usepackage{caption}
\usepackage{subcaption}
\usepackage{amsmath}
\usepackage{hyperref}  % For clickable links in PDF
\usepackage{cleveref}  % For smart references
\usepackage[acronym]{glossaries}

\mathchardef\mhyphen="2D

\newacronym{ros}{ROS}{Robot Operating System}
\newacronym{vrpn}{VRPN}{Virtual Reality Peripheral Network}
\newacronym{ugv}{UGV}{Unmanned Ground Vehicle}
\newacronym{anova}{ANOVA}{Analysis of variance}
\newacronym{udp}{UDP}{User Datagram Protocol}

\title{\LARGE \bf
Understanding Dynamic Human-Robot Proxemics in the Case of Four-Legged Canine-Inspired Robots
}
\author{Xiangmin Xu$^{1}$, Zhen Meng$^{1}$, Emma Li$^{1}$, Mohamed Khamis$^{1}$, Philip G. Zhao$^{2}$ and Robin Bretin$^{1}$% <-this % stops a space
\thanks{$^{1}$Xiangmin Xu, Zhen Meng, Emma Li, Robin Bretin, and Mohamed Khamis are with the School of Computing Science, University of Glasgow, Glasgow, G12 8QQ, UK. 
{\tt\small \{x.xu.1, r.bretin.1\}@research.gla.ac.uk, \{zhen.meng, liying.li, mohamed.khamis\}@glasgow.ac.uk}}%
\thanks{$^{2}$Philip G. Zhao is with the Department of Computer Science, University of Manchester, Manchester, M13 9PL, UK. 
        {\tt\small philip.zhao@manchester.ac.uk}}%
}

\begin{document}
\maketitle
\thispagestyle{empty}
\pagestyle{empty}
\begin{abstract}
% Recently, quadruped robots have been well developed with potential applications in different areas, such as care homes, hospitals and other social areas. 
% To ensure their integration in such social contexts, it is essential to understand people's proxemic preferences around such robots. 
% In this paper, we designed a human-quadruped-robot interaction study to investigate the effect of 1) different facing orientations and 2) the gaze of a moving robot on human proxemic distance. Specifically, our work covers both static and dynamic interaction scenarios. We found a statistically significant effect of both the robot's facing direction and its gaze on preferred personal distances. The distances humans established towards certain robot behavioral modes reflect their attitudes, thereby guiding the design of future autonomous robots.
The integration of humanoid and animal-shaped robots into specialized domains, such as healthcare, multi-terrain operations, and psychotherapy, necessitates a deep understanding of proxemics—the study of spatial behavior that governs effective human-robot interactions. Unlike traditional robots in manufacturing or logistics, these robots must navigate complex human environments where maintaining appropriate physical and psychological distances is crucial for seamless interaction. This study explores the application of proxemics in human-robot interactions, focusing specifically on quadruped robots, which present unique challenges and opportunities due to their lifelike movement and form. Utilizing a motion capture system, we examine how different interaction postures of a canine robot influence human participants' proxemic behavior in dynamic scenarios. By capturing and analyzing position and orientation data, this research aims to identify key factors that affect proxemic distances and inform the design of socially acceptable robots. The findings underscore the importance of adhering to human psychological and physical distancing norms in robot design, ensuring that autonomous systems can coexist harmoniously with humans. 
\end{abstract}

% \begin{keywords}
% Human-Robot Interaction, Physical Human-Robot Interaction, Social HRI, Safety in HRI
% \end{keywords}
% \keywords{Human-Robot Interaction}
%

\sloppy
% \vspace{-2mm}
\section{Introduction}
\label{sec:intro}

Humanoid and animal-shaped robots are poised for significant implementation in real-world applications, representing an evolution beyond the established roles of traditional robotic systems such as robotic arms and \glspl{ugv}. While these conventional robots have become integral to industries such as manufacturing, medical surgery, and logistics, humanoid and animal-shaped robots are being developed for more specialized applications, including healthcare \cite{esterwood2021systematic}, multi-terrain operations \cite{adarsh2018multi}, and psychotherapy \cite{shamsuddin2012initial}. These specific use cases often require robots to exhibit highly human-like or animal-like forms to simulate realistic interactions, or to possess advanced mobility features, such as robotic limbs, to navigate complex environments.

\begin{figure}
    \centering
    \includegraphics[width=0.45\textwidth]{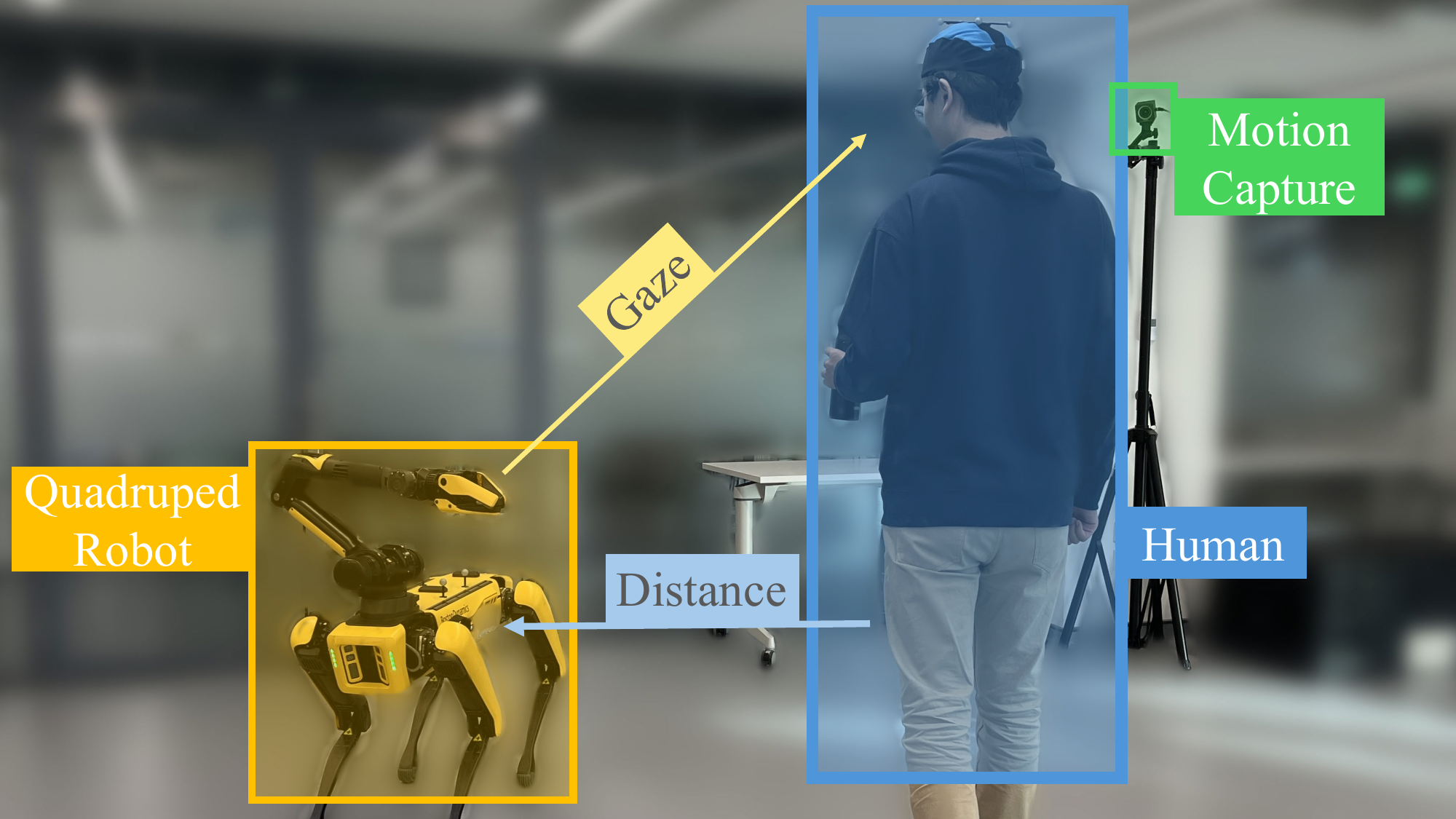}
    \caption{A participant interacting with our quadruped robot, with the trajectory recorded by a motion capture system.}
    \label{fig:teaser}
\end{figure}

A critical requirement for these robots, especially in scenarios where they interact closely with humans, is the ability to coexist and cooperate effectively within shared environments. To achieve this, these autonomous systems must be capable of aligning their behaviors with those of human teammates or adapting to human-centric environments \cite{mumm2011human}. Ensuring that robots exhibit appropriate proxemic behaviors—maintaining suitable physical distances and respecting social norms—is essential for their successful integration into human spaces.

At a minimum, robots must maintain a safe operational distance from humans, and human collaborators must be aware of and respect designated robot operation zones. These safety considerations are foundational, particularly for robots that pose a higher risk of harm, such as robotic arms or autonomous vehicles. Several researchers, including Safeea \cite{safeea2017minimum} and Camara \cite{camara2020space}, have proposed methodologies to secure these baseline interactions. However, while physical safety is a critical concern, it is insufficient on its own for the successful integration of robots into social settings. If robots fail to adhere to human norms regarding physical and psychological distancing, they may be perceived as threatening or disruptive \cite{mumm2011human}. The interpersonal distance maintained between humans is a key social signal that conveys levels of intimacy or discomfort, and the proxemic distance maintained by robots will similarly carry social implications. Therefore, to ensure that robots are perceived as friendly and reliable, it is essential to design them in accordance with societal psychological and physical norms.

Proxemic research, which has traditionally focused on human-human interactions, has increasingly been applied to human-robot scenarios. Takayama \cite{takayama2009influences} investigated the static distances preferred by humans in face-to-face human-robot interactions. Koay \cite{koay2014social} explored the influence of a robot’s orientation and subtle cues on its perceived likability in domestic settings. Hirota \cite{hirota2012mascot} examined the emotional responses of participants to robots moving at varying speeds and with different motion patterns, assessing the impact of mobility on proxemic outcomes. Bretin \cite{bretin2022co} conducted studies on personal distances maintained by participants when encountering drones in virtual reality, focusing on the influence of drone height.

Despite extensive research on traditional robotic systems, there has been limited investigation into quadruped robots, which represent a distinct category with superior vertical mobility compared to wheeled robots and a more lifelike appearance than humanoids or robotic arms. 

In this paper, we present a proxemic user study ($N$=32) involving a canine robot in a motion capture environment. The study involves a dynamic scenario where participants pass by the moving canine robot while engaged in an unrelated task, with their positions recorded in real-time by motion capture cameras for subsequent analysis. The objective of this experiment is to explore the impact of different interaction postures on participants' proxemic behavior, contributing to the broader understanding of human-robot interaction in dynamic environments.

\section{Related Work}
\label{sec:related}

\subsection{Proxemics}

\subsubsection{Human Proxemics}

The study of proxemic relationships, which explores the spatial behavior between individuals, has been a focus of research for many years \cite{mead2013automated, hall1960silent}. The authors in \cite{hall1966hidden} introduced the concept of proxemics, categorizing personal space into four distinct zones: intimate (6–18 inches) for close relationships, personal (1.5–4 feet) for family and close friends, social (4–12 feet) for acquaintances, and public (12–25 feet) for public settings. These zones help communicate varying levels of intimacy and relational distance during interactions.

The authors in \cite{aiello1987human} later expanded on this concept, suggesting that proxemic behavior serves two primary purposes: 1) to mitigate strong emotional responses triggered by close physical proximity \cite{blascovich1996biopsychosocial}, and 2) to maintain sufficient space to manage potential threats \cite{cavallin1980aggressiveness, bretin2022co}. Proxemic distances are not static; they change depending on the nature of the interaction and the entities involved. The interpersonal distances people naturally maintain reflect their attitudes toward others or objects in their environment, which in turn influences their behavior.

In modern research, the concept of proxemics has been extended to include interactions with robots \cite{takayama2009influences, walters2011a}, which are increasingly integrated into manufacturing and service industries \cite{pedersen2016robot}. While autonomous robots offer many benefits, they also raise concerns, such as privacy risks \cite{kaminski2016averting} and the potential for aggressive behavior \cite{walters2011a}.

\subsubsection{Human-Robot Proxemics}

As robots become more integrated into daily life, particularly in private spaces, understanding and adhering to human proxemic norms becomes crucial. This is especially true for robots used in home services \cite{xu2017developing, li2021design}. 
The authors in~\cite{mumm2011human} argue that robots must be designed to follow societal norms regarding physical and psychological distancing to interact effectively with humans. A key aspect of this is the manipulation of the robot gaze, which plays a critical role in psychological proxemics. Additionally, the static orientation, distance, and signaling of robots are significant factors that influence how humans perceive and interact with them \cite{koay2014social}. The authors in~\cite{eresha2013investigating} found that unwanted gaze during human-robot interactions can cause discomfort or distract the human participant, highlighting the importance of gaze in designing socially acceptable robots.

\subsection{Movement Capture in Proxemics}

Maintained distance is a critical metric for analyzing how people perceive objects they encounter and for assessing their overall sensory experience during interactions \cite{hall1995handbook, mead2011recognition}. A common approach in proxemics research is to measure the minimum distance between humans and robots over time and space as an indicator of personal distance \cite{bailenson2003interpersonal, bretin2022co, interrante2006distance}.

Various methods have been developed for studying indoor proxemics, including RFID \cite{cardenas2017proximithings}, virtual reality (VR) \cite{bretin2022co}, and wearable devices \cite{montanari2018measuring}. Although these technologies offer convenience and mobility, none match the precision of motion capture systems, which are widely used in the creation of state-of-the-art virtual reality experiences and 3D films due to their high precision, quick response, and high sampling rates \cite{chan2010virtual, menache2000understanding, bregler2007motion}. Research using motion capture systems in proxemics studies leverages the system’s capability to analyze participants' movements at all key body points \cite{kosinski2016fuzzy}, extracting features relevant to emotion or perception. For instance, the authors in~\cite{mead2013automated} used a Hidden Markov Model to recognize participants' behaviors based on motion data collected from a Microsoft Kinect camera. The authors in~\cite{jakobsen2013information} focused on position, velocity, and orientation to find efficient methods for information visualization in the context of proxemics.

\section{System Design}
\begin{figure}
    \centering
    \includegraphics[width=0.5\textwidth]{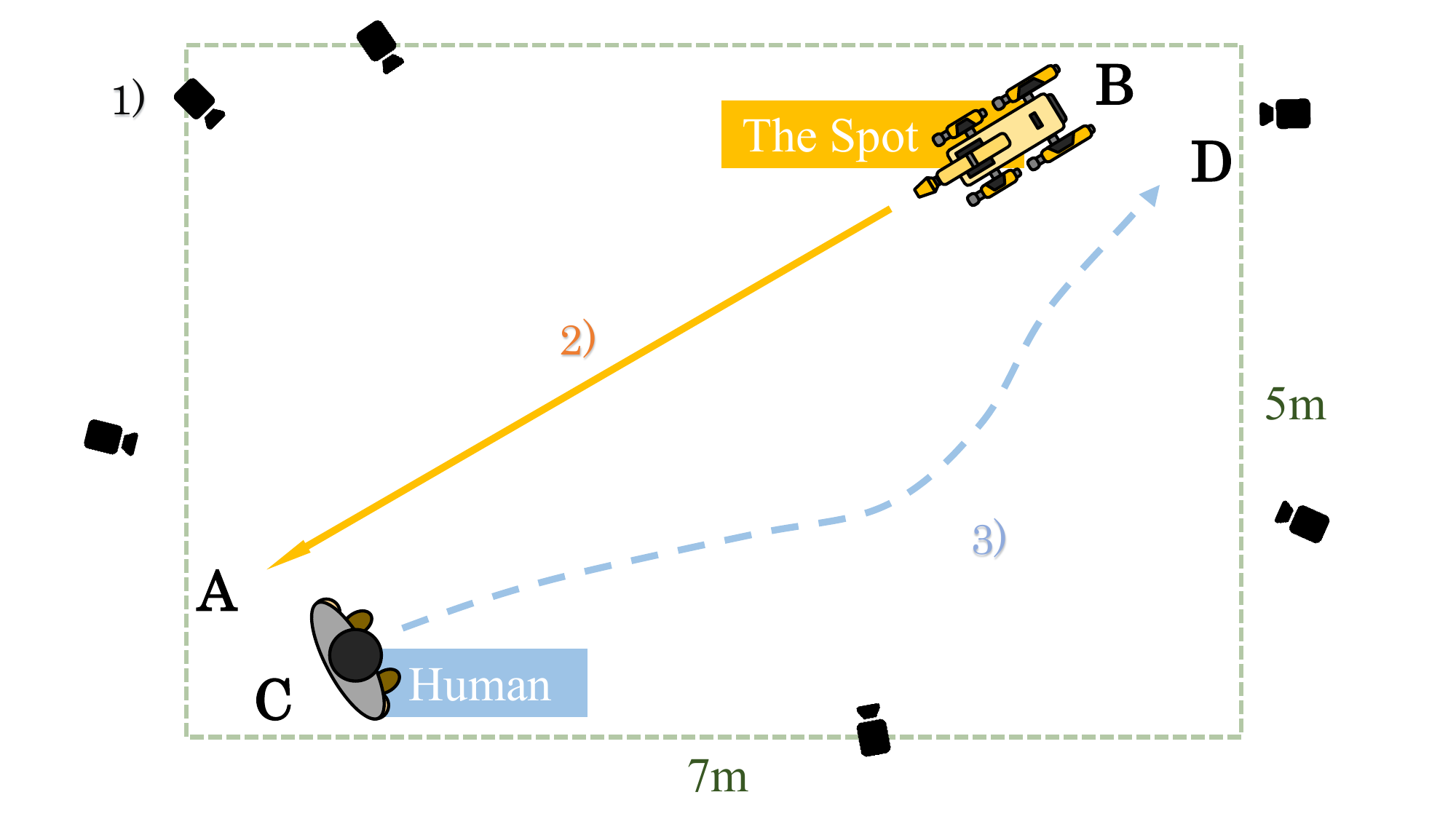}
    \caption{Design of the experiment: 1) OptiTrack Cameras, 2) Spot Trajectory. The Spot will always move from B to A if not still. Without gaze, in forward conditions, the Spot is facing A point; in sideways conditions, the Spot is vertical to line AB, facing the participant side, and in backward conditions, the Spot is facing B point. The head of the Spot will always face the participant in gaze conditions. 3) Participants' assumed trajectory. The participants will always face the dotted line direction at the beginning and move from C to D, and they can choose their own paths while moving.}
    \label{fig:setupa}
\end{figure}

% This experiment aims to investigate the influence of the canine robot's orientation and gaze on participants' proxemic behavior in a dynamic passing-by interaction scenario in the real world (compared to \acrshort{vr} methods). A fully still condition is added as a control condition. We use an OptiTrack \cite{opti} motion capture system to record the movement trajectory of both the robot and the participant at minimum interference. Bretin et al.~\cite{bretin2022co} argue that distancing is not a thoughtful decision, but rather an intuitive and straightforward process. 
% The method of stop distance \cite{fink2007obstacle, takayama2009influences, li2019comparing} is inappropriate in our case for its interaction pattern requires participants to think before they stop. 
% Instead of letting participants subjectively decide which distance they want to maintain with the canine robot, we integrate the process in a more subconscious way: the participants are asked to perform an unrelated task, in the middle of which the canine robot moves and passes by the participants. The personal distance is still the minimum mutual distance between humans and robots, which can be calculated from trajectories in a time series. This motion capture method has been successfully implemented by Marquardt et al.~\cite{marquardt2011proximity} in another proxemic study. 

As shown in Fig.~\ref{fig:setupa}, this work aims to investigate how the direction and gaze of a canine-inspired robot, such as Boston Dynamics' Spot~\cite{boston_dynamics}, influence participants' behavior during close-proximity interactions in real-world, dynamic passing scenarios. To achieve this, we designed a system using the OptiTrack motion capture system~\cite{optitrack}, which accurately records the movement trajectories of both the robot and the participants with minimal interference.

% \subsubsection{Motion Capture System} The lab is equipped with an OptiTrack motion capture system, which functions in the outside-in \cite{ribo2001new} tracking principle. Six motion cameras are mounted around the experiment zone to take 2D aligned pictures of passive markers on objects, according to the position of retroreflective markers on 2D frames to calculate the real world 3D marker position. The Motive software transfers certain shapes formed by markers into a rigid body, the markers were installed asymmetrically so that the orientation can be identified as in Figure \ref{fig:marker}. The rigid body coordinate system is left-handed, the same as the world coordinate. The rigid body parameters will be stored in OptiTrack configurations to make them recognizable in every experiment setup. With a sufficient frame rate, the system can capture the in-situ position of the marker rigid bodies in sight. The rigid bodies' information on positions and orientations is sampled at the rate of 100 Hz. The position information is then multi-casted in a local network with \acrfull{ros} \acrfull{vrpn} communication toolkit using the UDP protocol to guarantee communication speed.
\subsection{Motion Capture System}The motion capture component is built around the OptiTrack system, which is responsible for accurately capturing the movements of both the robot and the human participants.

\subsubsection{OptiTrack System} The system is equipped with six strategically placed cameras that track reflective markers attached to the robot and participants. These markers allow for precise 2D alignment and 3D position computation, enabling the accurate tracking of trajectories in real time.

\subsubsection{Data Processing and Transmission} The captured position and orientation data are processed by the OptiTrack system and broadcast locally using the \gls{ros} and \gls{vrpn} toolkits. Data transmission is performed over a local network using \gls{udp}, ensuring that the information is conveyed efficiently and reliably for real-time analysis.

\subsection{Canine-Inspired Robot} 
The robot component is central to the experiment, providing the movement and gaze behaviors that are the focus of the study.

\subsubsection{Locomotion Capabilities} The Boston Dynamics Spot robot is equipped with autonomous four-legged locomotion, allowing it to move in various directions—forward, backward, laterally, or stationary. These movement capabilities are essential for testing different interaction scenarios.

\subsubsection{Gaze Simulation} The robot is fitted with a mechanical arm that includes an RGBD camera. This camera enables the robot to simulate gaze behavior by orienting towards specific points in space, which is critical for studying the impact of gaze on human participants.

\subsubsection{Control Interface} The Spot robot’s movement and gaze behaviors are controlled through the Spot SDK, allowing precise configuration according to the experimental conditions.

\begin{figure}
    \centering
    \includegraphics[width=0.45\textwidth]{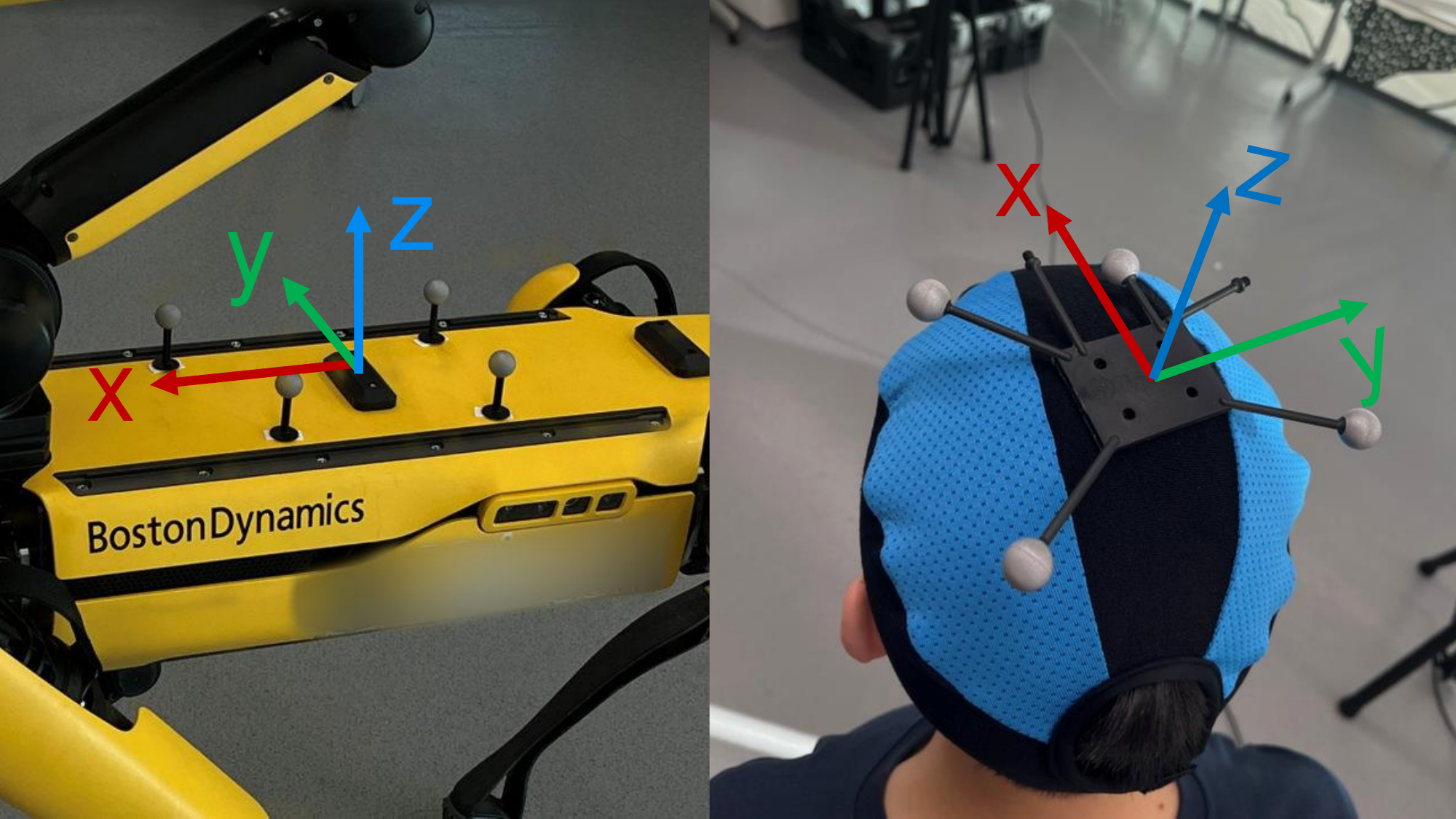}
    \caption{Rigid body markers on Spot and participants}
    \label{fig:marker}
\end{figure}

\begin{figure}
    \centering
    \includegraphics[width=0.47\textwidth]{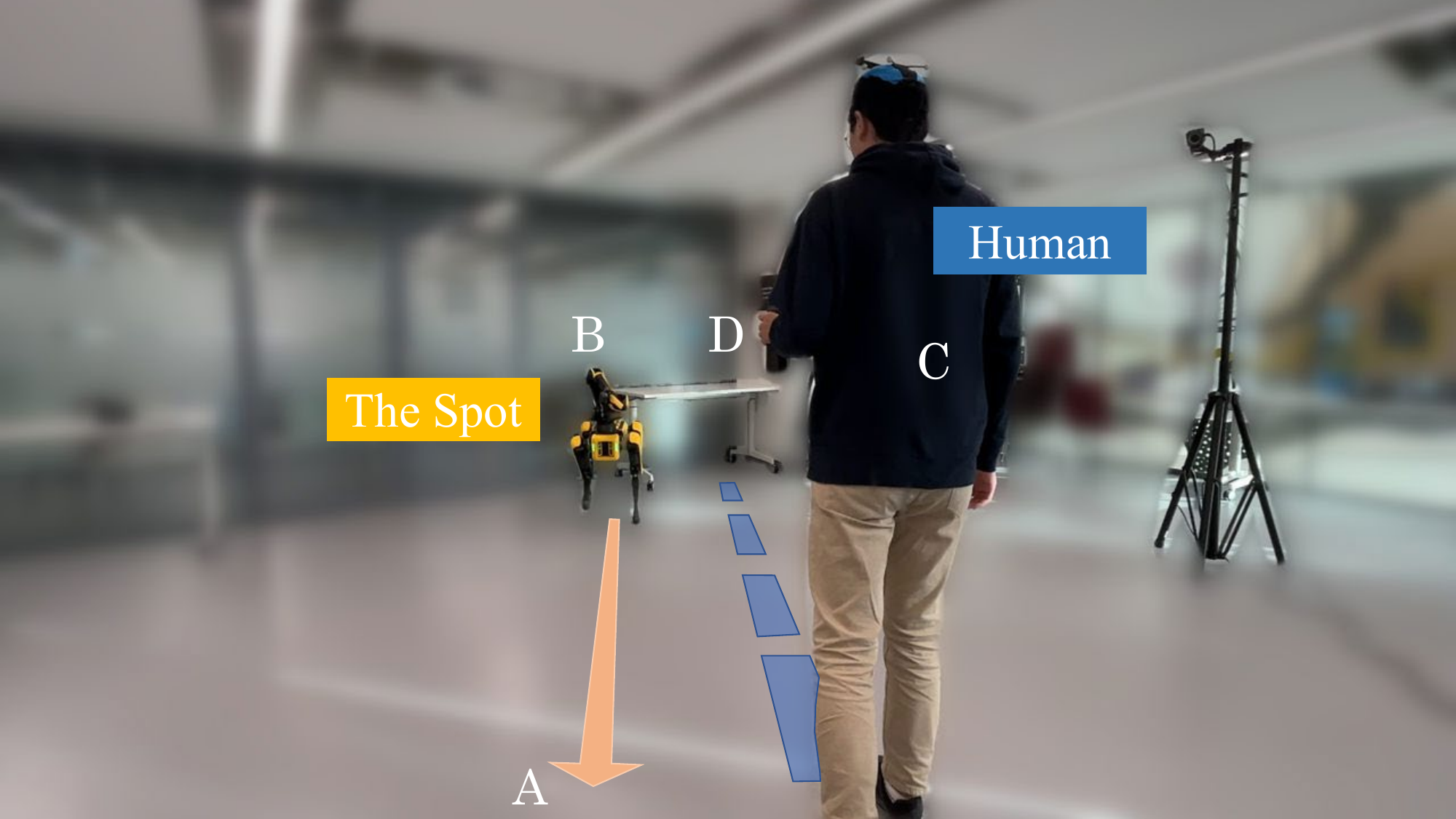}
    \caption{Experimental Setup.}
    \label{fig:setupb}
\end{figure}

\subsection{Human Participants}
The human participants are a vital component of the system, providing the interaction data necessary for analysis.

\subsubsection{Participant Tracking}Participants wear passive marker caps that allow the OptiTrack system to capture their movement trajectories as they perform predefined tasks within the experimental area.

\subsubsection{Task Execution}During the experiment, participants were instructed to carry out a specific task: transporting an object from one point to another within the experimental area. Their natural movement while completing this task was critical for analyzing how they interacted with the robot under various conditions. The study employed a 2x4 within-subjects design, where each participant experienced all experimental conditions. The experiment included two independent variables:

\begin{itemize}
    \item \textbf{Movement} (four conditions):
    \begin{enumerate}
        \item \textbf{Moving Forward (Baseline):} Spot moves from point B to point A while facing the participant.
        \item \textbf{Moving Sideways:} Spot moves laterally from point B to point A with its head oriented towards the participant’s pathway.
        \item \textbf{Moving Backward:} Spot moves from point B to point A with its back facing the participant.
        \item \textbf{Stationary:} Spot remains stationary, equidistant from points A and B.
    \end{enumerate}
    \item \textbf{Gaze} (two conditions):
    \begin{enumerate}
        \item \textbf{Gaze On:} Spot gazes directly at the participant.
        \item \textbf{Gaze Off:} Spot does not gaze at the participant; instead, its head is aligned with its body.
    \end{enumerate}
\end{itemize}

% Task Execution: During the experiment, participants are instructed to carry out a specific task, carrying an object from one point to another. Their natural movement while completing the task is crucial for understanding how they interact with the robot under different conditions. The study follows a 2x4 within-subjects design, in which every participant performed all conditions. There were two independent variables:
% \begin{itemize}
%     \item [] {\textbf{Movement} with four conditions: 
%     \begin{enumerate}
%     \item Moving forward (Baseline): Spot is moving from point B to point A while facing the participant,
%     \item Moving sideways: Spot is moving sideways from point B to point A with its head facing the pathway of the participant, 
%     \item Moving backward: Spot is moving from point B to point A with its back facing the participant. 
%     \item Stationery: Spot is stationary at equal distances from A and B,
%     \end{enumerate}}
%     \item [] {\textbf{Gaze} with two conditions: a) Gaze on: Spot is gazing at the participant, and b) Gaze off: it does not gaze, but its head is rather aligned with its body.}
% \end{itemize}

\begin{figure}
    \centering
    \text{(a)}
    \includegraphics[width=0.5\textwidth]{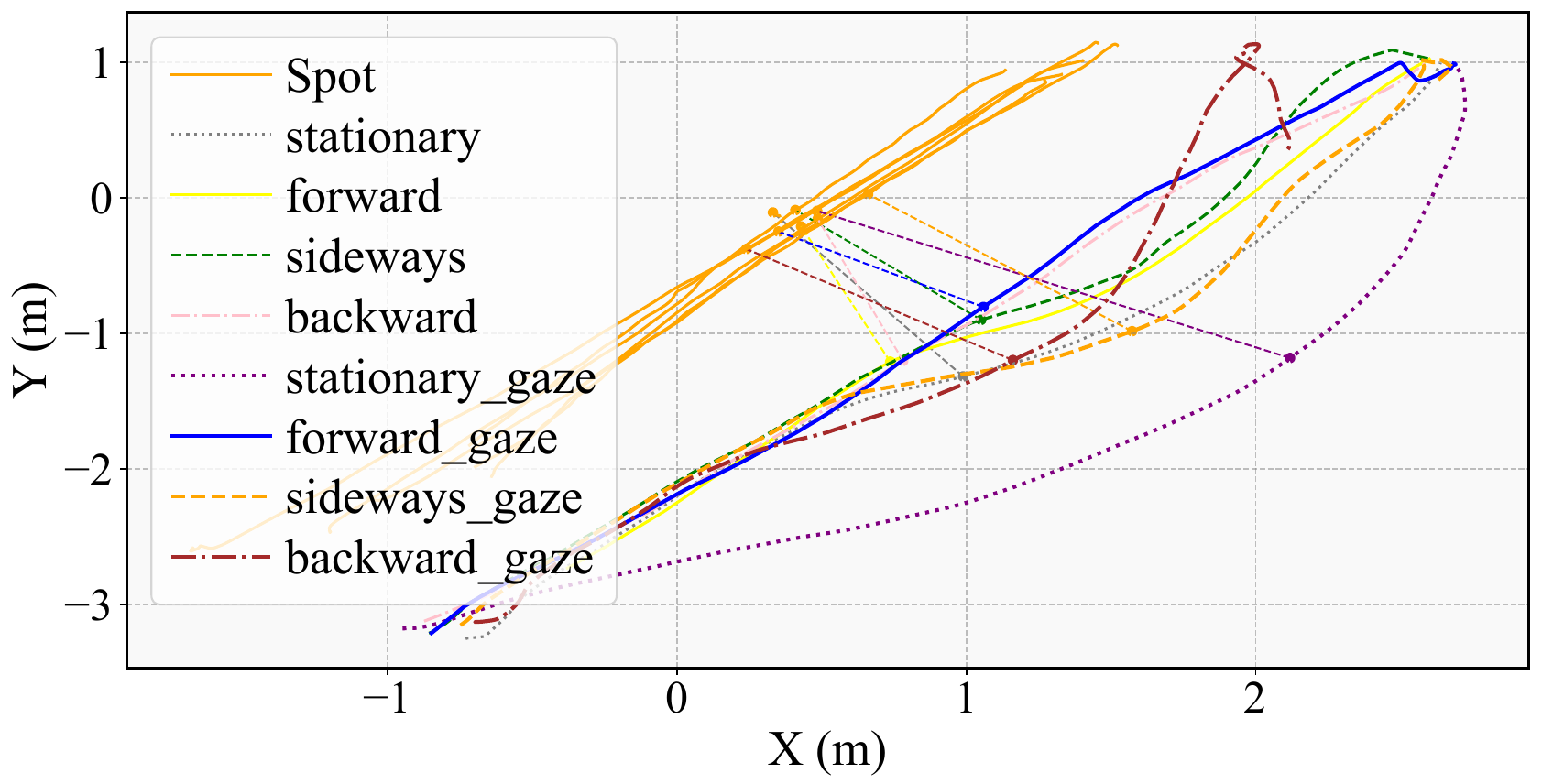}
    % \centering
    % \text{(b)}
    % \includegraphics[width=0.5\textwidth]{figs/bar.pdf}
    \caption{(a): Trajectories of a single participant. The orange fixed trajectory for Spot, other colored trajectories for humans, and the time-synchronized minimum distances are labeled as dotted lines in the plot. 
    % (b): Distance in bar plot corresponding to trajectories in (a).
    }
    \label{fig:traj}
\end{figure}

\subsection{Data Collection}

We recruited 32 participants (17 males, 15 females), with ages ranging from 19 to 37 years ($M = 26, SD = 3.57$). Their experience with robots varied. This study was approved by our ethics committee. 
% (ethics application \#XXXXX\footnote{Redacted for anonymous review}). 
Each participant underwent 2$\times$4 = 8 different conditions, with the order of conditions counterbalanced using an 8$\times$8 Latin square. Participants were not informed of the sequence of cases.

As shown in Fig.~\ref{fig:setupa}, participants were instructed to retrieve a cup from the table at position A. Participants were informed that their task was to bring the cup to the table at position D. The Spot robot's movement was updated in accordance with each condition as the participant performed their task. 
Firstly, participants started from position C and moved toward the table at position D to deliver the yellow cup, during which the trajectories of the participant and the Spot were recorded by the motion capture system. In non-stationary conditions, the Spot robot began moving from B to A as soon as the participant started moving from C to D. Participants were instructed to walk at any preferred speeds or trajectories. During the process, the OptiTrack system recorded the 3D pose data of both the participants and the Spot. After eight repetitions of the same experiment for each individual, the experimental collection process ended. After the experiment, participants completed a questionnaire to collect demographic information and details about their robot experience.

\begin{figure*}
    \centering
    \includegraphics[width=\linewidth]{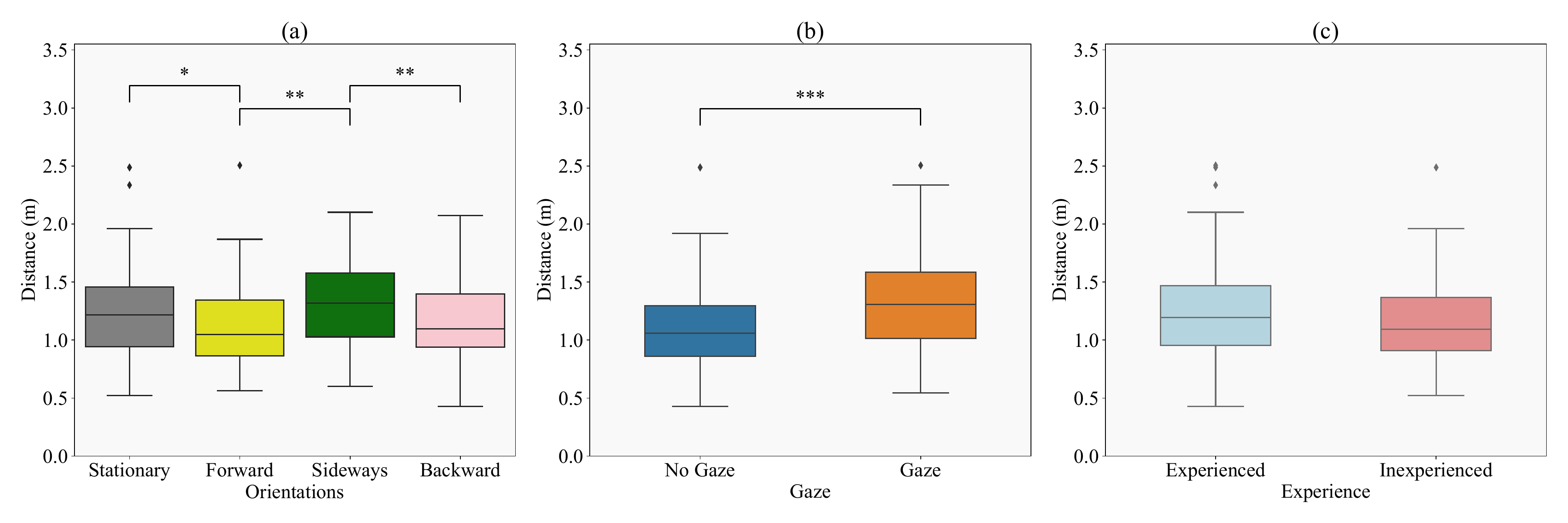}
    \caption{Distance in different condition groups:
    (a) The personal distance in four different orientation conditions.
    (b) The personal distance in two different gaze conditions
    (c) The personal distance in two different experience conditions.
    }
    \label{fig:gaze_anova}
\end{figure*}

\section{Evaluation}
\subsection{2D Trajectories Visualization}

The 2D trajectories of participants across different scenarios are shown in Fig. \ref{fig:traj}. These trajectories were obtained by projecting the original 3D trajectory data onto the xy-plane. We disregarded the Z-axis distance due to 1) the height variations among participants and 2) the natural vertical oscillations that occur during walking. This approach was chosen to standardize the comparison of movement patterns across different individuals and scenarios, ensuring that the analysis focuses on horizontal movement dynamics, which are more relevant to personal space and proxemics in the context of human-robot interaction.

\subsection{Hypotheses and Results} 
% We performed a student t-test with the factors of robot orientation, gaze involvement, and user experience
% \begin{align}
% t = \frac{\bar{d}_1 - \bar{d}_2}{\sqrt{\frac{s_1^2}{n_1} + \frac{s_2^2}{n_2}}},
% \end{align}
% where $\bar{d}_1$ and $\bar{d}_2$ are the mean values of distance regarding different scenarios, $s$ is the standard deviation and $n$ is the number of participants.

Based on the results we observed from the above distances, we made several assumptions and performed a two-way $2 * 4$ \gls{anova} with two within-participant factors: the robot gaze factor of two levels (with/without), and the robot orientation of four levels (stationary, forward, sideways and backward). The dependent variable is the time-synchronized minimum participant-robot distance for each set of conditions. We verified the data on the normality assumption (Shapiro-Wilk test, $p > 0.05$) and the homogeneity of variances (Levene’s test, $p > 0.05$). The \gls{anova} test revealed a significant main effect of the robot orientation on distance ($F(3, 256) = 7.53, p = 0.00008, partial~\eta^2 = 0.091$). Gaze also showed a significant main effect ($F(1, 256) = 21.89, p = 0.000005, partial~\eta^2 = 0.089$). However, the interaction effect between orientation and gaze was not significant, ($F(3, 256) = 0.69, p = 0.56, partial~\eta^2 = 0.009$).

\textbf{H1: The personal distance in the sideways condition is higher than that in the forward condition.}

This hypothesis stems from the expectation that participants would maintain greater distances from the robot when it moves sideways. The sideways orientation is unique as it introduces an unconventional movement pattern where the robot’s head is oriented perpendicular to its trajectory, moving orthogonally to its facing direction. This orientation might increase participants' perceived unpredictability of the robot’s movements, leading them to maintain a larger personal distance as a precaution.

The results strongly support this hypothesis. As shown in Figure \ref{fig:gaze_anova}(a), participants did indeed maintain a greater distance in the sideways movement condition ($M = 1.35, SD = 0.44$) compared to the forward movement condition ($M = 1.10, SD = 0.35$). The two-way \gls{anova} has proved the significant effect of robot orientation on personal distances. To further examine the significant differences among various orientations, we conducted a post hoc analysis using Tukey's Honest Significant Difference (HSD) among the four orientation conditions. The results revealed that the personal distances in the forward orientation condition are significantly longer than in the sideways orientation condition ($\Delta = 0.2605, 95\%~CI~[0.0081, 0.4399], p = 0.0012 < 0.05$). So we have the conclusion that the sideways orientation indeed causes participants to maintain a greater distance, thus confirming H1.

\textbf{H2: The personal distance in the backward condition is higher than that in the forward condition.}

This hypothesis posits that participants would maintain longer distances from the robot when it moves backward, compared to when it moves forward. The rationale behind this hypothesis is the assumption that backward movement might induce uncertainty or discomfort, as it deviates from the more predictable forward movement commonly observed in vehicles and other robots.

However, the results did not support this hypothesis. In the backward movement condition ($M = 1.10, SD = 0.33$), participants did not maintain a significantly different personal distance compared to the forward movement condition ($M = 1.10, SD = 0.35$), as shown in Figure \ref{fig:gaze_anova}(a). The Tukey's HSD test ($\Delta =  -0.0105, 95\%~CI~[-0.19,  0.169], p = 0.9988$) also revealed no significant difference in the mean minimum distance between the backward and forward orientation conditions. These findings suggest that contrary to our expectations, backward movement does not significantly increase participants' personal distance, leading to the rejection of H2. This outcome could be attributed to the familiarity of backward movement in vehicles, reducing its perceived unpredictability compared to the sideways orientation.

\textbf{H3: The personal distance will increase when gazing is induced.}

This hypothesis is based on prior research \cite{mumm2011human}, which suggests that gaze significantly affects the personal physical distance that humans maintain from robots. Specifically, when a robot engages in gazing behavior, participants are likely to increase their personal distance due to the heightened sense of being observed, which can evoke discomfort or a sense of intrusion.

The results provide strong evidence in support of this hypothesis. The mean minimum distance with the gaze ($M = 1.32, SD = 0.37$) increased significantly when compared with the no-gaze ($M = 1.10, SD = 0.37$) condition. Besides the ANOVA test that confirmed this hypothesis, we have also performed a t-test to examine the effect of robot gaze on human-robot interaction distance. As illustrated in Figure \ref{fig:gaze_anova}(b), The analysis revealed a significant difference between the gaze and no-gaze conditions ($t$-$statistic = 4.50, p = 0.00001$). These findings suggest that robot gaze is an important factor in the proxemic relationships between humans and quadruped robots. Participants tended to position themselves farther from the robot when it displayed gaze behavior, potentially indicating increased social presence or perceived autonomy of the robot.

\textbf{H4: The participants' robot experience will change their personal distances.}

This hypothesis explores whether participants with experience in robotics, either through professional engagement or personal interest, maintain different personal distances compared to those without such experience. Previous studies \cite {takayama2009influences} and \cite{mumm2011human}, suggest that individuals with more experience or familiarity with robots might feel more comfortable and thus maintain shorter distances. The robotic experience is a between-participant factor and is acquired from post-experiment questionnaires. 

Our results, however, offer a different perspective. The experienced group ($N = 12, M = 1.24, SD = 0.38$) maintained a slightly longer mean distance than the inexperienced group ($N = 20, M = 1.16, SD = 0.36$), as shown in Fig. \ref{fig:gaze_anova}(c). Interestingly, this observation contradicts the hypothesis, suggesting that more experienced participants might tend to be more cautious toward the quadruped robot, possibly due to their deeper understanding of robotic behaviors and potential risks. In addition, we performed a t-test to see the effect of experience on personal distances, and the result indicated that this difference was not significant ($t$-$statistic = -1.56, p = 0.12$). This lack of significance may be attributed to the small sample size of experienced participants ($N = 12$), which limits the power of the statistical analysis. While the results do not provide strong evidence to support H4, they do highlight an interesting trend that warrants further investigation with a larger sample size. The finding that experienced participants may maintain greater distances could be influenced by their knowledge and understanding of the potential unpredictability or operational nuances of robots like the Spot.

In summary, our evaluation confirmed H1 and H3, providing clear evidence that sideways orientation and gaze significantly increase personal distances in human-robot interactions. H2, which hypothesized an increase in distance with backward movement, was not supported by the data, and H4, regarding the impact of robotic experience, revealed intriguing but inconclusive results. These findings contribute to the understanding of how different robot behaviors and participant characteristics influence proxemic dynamics, offering valuable insights for designing autonomous systems that interact closely with humans.

\section{Discussion}

\subsection{The Stationary Position} Initially, we hypothesized that moving conditions would result in longer personal distances than stationary scenarios. However, our findings contradicted this assumption, with the stationary condition ($M = 1.28, SD = 0.36$) exhibiting higher personal distances than the forward-moving condition ($M = 1.10, SD = 0.35$). Tukey's HSD test confirmed the significance of this difference ($\Delta = 0.1833, 95\%~CI~[0.0038,0.3627], p = 0.0433 < 0.05$). Participants reported in post-experiment questionnaires that the stationary Spot appeared suspicious, raising concerns about potential sudden movements. This reaction likely stems from the robot's lifelike, ready-to-move posture, which contrasts with traditional wheeled robots that exhibit no visible difference when powered off. The uncertainty surrounding the robot's stillness, combined with its standing posture, likely heightened participants' caution, leading to greater personal distances.

\subsection{The Robotic Experience} We hypothesized that participants with more experience in robotics would maintain shorter distances from the canine robot. Contrary to this, the data suggested that experienced participants maintained slightly longer distances, though not statistically significant. Questionnaire responses indicated that experienced participants were more cautious, possibly due to awareness of potential bugs or algorithmic failures. In contrast, participants with little to no robotic experience, driven by curiosity and the robot's friendly appearance, tended to approach more closely. This suggests that experience may induce a more cautious approach due to perceived risks, while inexperience fosters curiosity-driven proximity.

\subsection{Limitations} While robotic gaze was a central focus, this study did not account for the participant's attention to the robot. Some participants did not consistently engage with the robot visually, particularly during crossings. Additionally, the impact of environmental noise, such as the Spot's fan and motor sounds, was not considered. These noises may have influenced participants' perceptions, possibly making the robot seem more intimidating. Future work should examine the role of human attention and environmental factors, such as noise, in shaping human-robot proxemic interactions.

\section{Conclusion and Future Work} This study explored human-robot proxemics using a canine-inspired quadrupedal robot, focusing on the effects of movement orientation, gaze, and robotic experience on personal distance. We found that the shortest distances occurred when the robot moved in its facing direction without gazing at the participants. Sideways movement and gaze led to increased distances due to heightened insecurity and uncertainty about the robot's behavior. Additionally, experience with robotics tended to increase personal distances, likely due to greater awareness of potential risks. Our findings suggest that robots moving in their facing direction with minimal gaze can be perceived as less threatening and more approachable in human social spaces. Future research should investigate the impact of the robot's posture, noise, and the attention level of participants on proxemic behavior to deepen our understanding of human-robot interactions.

\section{Acknowledgement}
This work was supported by the UKRI Centre for Doctoral Training in Socially Intelligent Artificial Agents, Grant Number EP/S02266X/1.

\bibliographystyle{IEEEtran}
\bibliography{bib}

\end{document}